\documentclass[draft]{agujournal2019}
\usepackage{url} 
\usepackage{lineno}
\usepackage{booktabs}
\usepackage{graphicx}
\usepackage{subcaption}
\usepackage{float}

\usepackage{soul}
\draftfalse

\journalname{Journal of Advances in Modeling Earth Systems (JAMES)}

\begin{document}

%
%


\title{Training neural mapping schemes for satellite altimetry with simulation data}
\authors{Q. Febvre\affil{1,4}, J. Le Sommer\affil{2}, C. Ubelmann\affil{3}, R. Fablet\affil{1,4}}

\affiliation{1}{IMT Atlantique, Lab-STICC, Brest, France}
\affiliation{2}{Université Grenoble Alpes, CNRS, IRD, Grenoble, France}
\affiliation{3}{Datlas, Grenoble, France}
\affiliation{4}{INRIA Team ODYSSEY}
\correspondingauthor{Quentin Febvre}{quentin.febvre@imt-atlantique.fr}

\begin{keypoints}

\item We propose to train neural mapping schemes for real altimeter data from ocean simulation data.
\item  The trained neural schemes significantly outperform the operational mapping of real altimetry data for a Gulf Stream case-study.
\item More realistic simulation datasets improve the performance of the trained neural mapping with a 20\% improvement in the spatial  scales.
\end{keypoints}

\begin{abstract}
    Satellite altimetry combined with data assimilation and optimal interpolation schemes have deeply renewed our ability to monitor sea surface dynamics. Recently, deep learning (DL) schemes have emerged as appealing solutions to address space-time interpolation problems. The scarcity of real altimetry dataset, in terms of space-time coverage of the sea surface, however impedes the training of state-of-the-art neural schemes on real-world case-studies. Here, we leverage both simulations of ocean dynamics and satellite altimeters to train simulation-based neural mapping schemes for the sea surface height and demonstrate their performance for real altimetry datasets. We analyze further how the ocean simulation dataset used during the training phase impacts this performance. This experimental analysis covers both the resolution from eddy-present configurations to eddy-rich ones, forced simulations vs. reanalyses using data assimilation and tide-free vs. tide-resolving simulations. Our benchmarking framework focuses on a Gulf Stream region for a realistic 5-altimeter constellation using NEMO ocean simulations and 4DVarNet mapping schemes. All simulation-based 4DVarNets outperform the operational observation-driven and reanalysis products, namely DUACS and GLORYS. The more realistic the ocean simulation dataset used during the training phase, the better the mapping. The best 4DVarNet mapping was trained from an eddy-rich and tide-free simulation datasets. It improves the resolved longitudinal scale from 151 kilometers for DUACS and 241 kilometers for GLORYS to 98 kilometers and reduces the root mean squared error (RMSE) by 23\% and 61\%. These results open research avenues for new synergies between ocean modelling and ocean observation using learning-based approaches.

\end{abstract}

\section*{Plain Language Summary}

For an artificial intelligence (AI) to learn, one need to describe a task using data and an evaluation procedure. 
Here we aim at constructing images related to the ocean surface currents. The satellite data we use provide images of the ocean surface with a lot of missing data (around 95\% of missing pixels for a given day), and we aim at finding the values of the missing pixels.
Because we don't know the full image, it is challenging to train an AI on this task using only the satellite data.
However, today's physical knowledge makes it possible to numerically simulate oceans on big computers. For these simulated oceans, we have access to the gap-free image, so we can train AI models by first hiding some pixels and checking if the model fill the gaps with the correct values.
Here, we explore under which conditions AIs trained on simulated oceans are useful for the real ocean.
We show that today's simulated oceans work well for training an AI on this task and that training on more realistic simulated oceans improve the performance of the AI!

\section{Introduction}

  Satellite altimeters have brought a great leap forward in the observation of sea surface height on a global scale since the 80's.

  Altimetry data have greatly contributed to the monitoring and understanding of key processes such as the sea-level rise and the role of mesoscale dynamics.
    The scarce and irregular sampling of the measurements presents a challenge for training deep neural networks.
  The retrieval of mesoscale-to-submesoscale sea surface dynamics for horizontal scales smaller than 150 km however remains a challenge for operational systems based on optimal interpolation \cite{taburetDUACSDT2018252019} and data assimilation \cite{jean-michelCopernicusGlobal122021} schemes. This has motivated a wealth of research to develop novel mapping schemes \cite{ballarottaDynamicMappingAlongTrack2020,ubelmannReconstructingOceanSurface2021,guillouMappingAltimetryForthcoming2021}.

  In this context, data-driven and learning-based approaches \cite{alveraazcarateReconstructionIncompleteOceanographic2005,barthDINCAEMultivariateConvolutional2022,lguensatAnalogDataAssimilation2017,fabletENDTOENDPHYSICSINFORMEDREPRESENTATION2021,martinSynthesizingSeaSurface2023} appear as appealing alternatives to make the most of the available observation and simulation datasets. Especially, Observing System Simulation Experiments (OSSE) have stressed the potential of neural schemes trained through supervised learning for the mapping of satellite-derived altimetry data \cite{fabletENDTOENDPHYSICSINFORMEDREPRESENTATION2021,beauchamp4DVarNetSSHEndtoendLearning2023}. 
  Their applicability to real datasets has yet to be assessed and recent studies have rather explored learning strategies from real gappy multi-year altimetry datasets \cite{martinSynthesizingSeaSurface2023}. Despite promising results, schemes trained with unsupervised strategies do not reach the relative improvement of the operational processing suggested by OSSE-based studies.
  
Here, we go beyond using OSSEs as benchmarking-only testbeds. We explore their use for the training of neural mapping schemes and address the space-time interpolation of real satellite altimetry observations. Through numerical experiments on a Gulf Stream case-study with a 5-nadir altimeter constellation, our main contributions are three-fold. We demonstrate the relevance of the simulation-based learning of neural mapping schemes and their generalization performance for real nadir altimetry data. We benchmark the proposed approach with state-of-the-art operational products as well as neural schemes trained from real altimetry datasets. We also assess how the characteristics of the training datasets, especially in terms of resolved ocean processes, drives the mapping performance.
To ensure the reproducibility of our results, our code is made available through an open source license along with the considered datasets and the trained models \cite{febvreCodeDataRelease2023a}.


The content of this paper is organized as follows. Section \ref{sec:background} offers background information on related work, Section \ref{sec:method} presents our method, Section \ref{sec:results} reports our numerical experiments, and Section \ref{sec:discussion} elaborates on our main contributions.

\section{Background}
\label{sec:background}
\subsection{Gridded satellite altimetry products}
\label{ssec:interpolation}
The ability to produce gridded maps from scattered along-track nadir altimeter measurements of sea surface height is key to the exploitation of altimeter data in operational services and science studies \cite{abdallaAltimetryFutureBuilding2021}.
As detailed below, we can distinguish three categories of approaches to produce such maps: reanalysis products \cite{jean-michelCopernicusGlobal122021} using data assimilation schemes, observation-based products \cite{taburetDUACSDT2018252019} and learning-based approaches \cite{fabletENDTOENDPHYSICSINFORMEDREPRESENTATION2021}.

Reanalysis products such as the GLORYS12 reanalysis \cite{jean-michelCopernicusGlobal122021} leverage the full expressiveness of state-of-the-art ocean models. They aim at retrieving ocean state trajectories close to observed quantities through data assimilation methods including among others Kalman filters and variational schemes \cite{carrassiDataAssimilationGeosciences2018}. Such reanalyses usually exploit satellite-derived and in situ data sources. For instance, GLORYS12 reanalysis assimilates satellite altimetry data, but also satellite-derived observations of the sea surface temperature, sea-ice concentration as well as in situ ARGO data  \cite{wongArgoData19992020}.

The second category involves observation-based products. In contrast to reanalyses, they only rely on altimetry data and address a space-time interpolation problem. They usually rely on simplifying assumptions on sea surface dynamics. In this category, optimal-interpolation-based product DUACS (Data Unification and Altimeter Combination System) \cite{taburetDUACSDT2018252019} exploits a covariance-based prior, while recent studies involve quasi-geostrophic dynamics to guide the interpolation scheme \cite{guillouMappingAltimetryForthcoming2021,ballarottaDynamicMappingAlongTrack2020}.

Data-driven and learning-based approaches form a third category of SSH mapping schemes. 
Similarly to observation-based methods, they are framed as interpolation schemes.
Especially deep learning schemes have gained some attention. Recent studies have explored different neural architectures both for real and OSSE altimetry datasets \cite{archambaultMultimodalUnsupervisedSpatioTemporal2023,beauchampDatadrivenLearningbasedInterpolations2021,martinSynthesizingSeaSurface2023}. These studies investigate both different training strategies as well as different neural architectures from off-the-shelf computer vision ones such as convolutional LSTMs and UNets \cite{ronnebergerUNetConvolutionalNetworks2015} to data-assimilation-inspired ones \cite{beauchampDatadrivenLearningbasedInterpolations2021, fabletLearningVariationalData2021}.

\subsection{Ocean Modeling and OSSE}
\label{ssec:oceanmodeling}
Advances in modeling and simulating ocean physics have largely contributed to a better understanding of the processes involved in the earth system and to the development of operational oceanography \cite{bernardImpactPartialSteps2006,ajayiSpatialTemporalVariability2020}. 
High-resolution simulations used in Observing System Simulation Experiments (OSSE) also provide a great test-bed for the design and evaluation of new of ocean observation systems \cite{benkiranAssessingImpactAssimilation2021}.
The availability of numerical model outputs enables the computation of interpretable metrics directly on the quantities of interest. This avoids challenges met when working solely with observation data that may be incomplete, noisy or indirectly related to the desired quantity.
For example, in the case of the recently launched SWOT mission, OSSEs combined ocean and instrument simulations to address calibration issues and interpolation performance for SWOT altimetry data  \cite{dibarboureDataDrivenCalibrationAlgorithm2022}.
Such OSSEs have also promoted novel developments for the interpolation of satellite altimetry such as the BFN-QG and 4DVarNet schemes \cite{guillouMappingAltimetryForthcoming2021,beauchamp4DVarNetSSHEndtoendLearning2023}. 

In OSSE settings, we can train learning-based mapping schemes in a supervised manner using 
model outputs as the "ground truth" during the training phase.
Nonetheless, these training methods cannot be straightforwardly applied to Observing System Experiments (OSEs) due to a lack of comprehensive groundtruthed observation datasets.
Applied machine learning practitioners often grapple with insufficient amount of labelled data during the training of supervised learning schemes, as the collection of large annotated datasets for a specific task can be costly or unattainable.
Proposed solutions includes the exploitation of large existing datasets (such as ImageNet \citeA{dengImageNetLargescaleHierarchical2009}) to train general purpose models (like \citeA {heDeepResidualLearning2016}). Another approach involves the generation of synthetic datasets to facilitate the creation of groundtruthed samples \cite{gomezgonzalezVIPVortexImage2017,dosovitskiyFlowNetLearningOptical2015}. OSSEs, which combine ocean model outputs and observing system simulators \cite{boukabaraCommunityGlobalObserving2018}, can deliver such large synthetic groundtruthed datasets. We propose to investigate how OSSE-based training strategies apply to the analysis of real satellite altimetry datasets. Recent results of SSH super-resolution model trained on simulation datasets and evaluated on real ones \cite{buongiornonardelliSuperResolvingOceanDynamics2022} support the relevance of such strategies.


\subsection{Physics-aware deep-learning}
\label{ssec:deeplearning}
In the last decades, DL advances combined with the rise in computational resources and amount of data have shown the power of extracting knowledge from data in domains ranging from computer vision to language processing \cite{lecunDeepLearning2015}. 
Yet, despite to the universality of DL architectures \cite{hornikMultilayerFeedforwardNetworks1989}, a central challenge persists in learning from data: the generalization performance beyond the distribution of the training data. To tackle this problem, the literature includes a variety of strategies such as data augmentation \cite{shortenSurveyImageData2019} and regularization techniques, including dropout layers \cite{srivastavaDropoutSimpleWay2014} and weight decay schemes \cite{krizhevskyImageNetClassificationDeep2012}. This is of critical importance for physical systems, where models trained on past data will be challenged when the system evolves and reaches dynamics absent from the training data. We can see evidence of this shortcoming in the instability challenges faced by neural closures for climate models \cite{brenowitzInterpretingStabilizingMachineLearning2020}. 

There have been a variety of approaches to harness physical priors within learning schemes to address this issue. Some injects trainable components in classical integration schemes of physical models such as \citeA{yinAugmentingPhysicalModels2021b}. Others leverage physical priors within their learning setups which can been used in the training objective \cite{raissiPhysicsinformedNeuralNetworks2019,greydanusHamiltonianNeuralNetworks2019}, as well as in the architecture \cite{li2020fourier,Wang2020TF}. 
However most of these works have focused on relatively simple physical models and it remains challenging to combine current state-of-the-art ocean models with such methods. Obstacles include the complexity and cost of running the physical models, the differences in programming tools and the computing infrastructures used in each domain, as well as the availability of automatic differentiation tools for state-of-the-art ocean models.

The proposed simulation-based training strategy offers another way to benefit from the advances in high-resolution ocean modeling in the design of deep neural models for ocean reanalysis problems.

\section{Method}
\label{sec:method}

\subsection{Overview}
\label{ssec:overview}

We designate our approach as "simulation-based", it consists in leveraging ocean models and simulations of observing systems to design supervised training environments.
In this section, we describe the proposed method for assessing the potential of simulation-based neural schemes for the mapping real altimetry tracks. We describe the architecture considered in our study, as well as the different datasets used for training purposes. We also detail our simulation-based training setup and the proposed evaluation framework on real altimetry.  

\begin{figure}[ht]
    \centering
    \includegraphics[width=\textwidth]{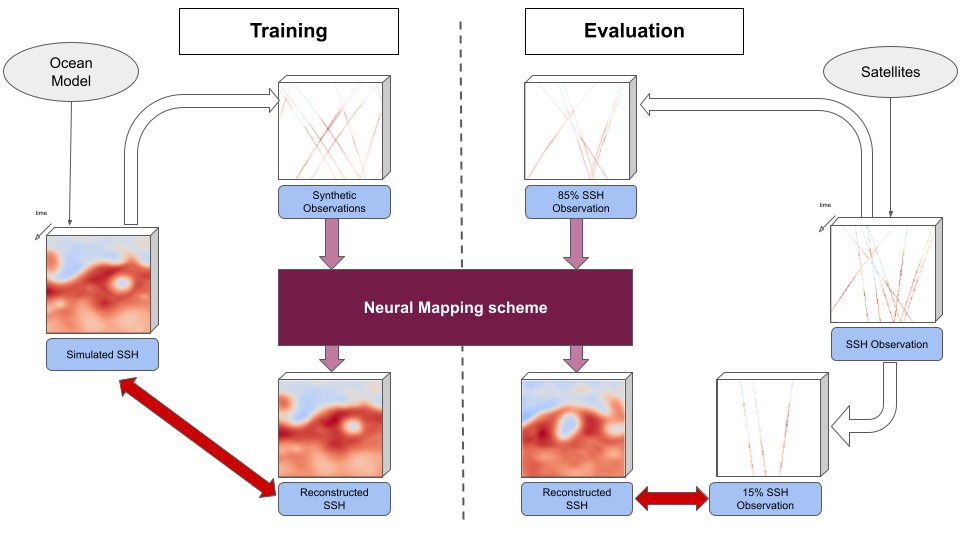}
    \caption{\textbf{Overview of the experimental setup}. On the left side we display the simulation-based training strategy based on an ocean simulation which will be used for 1) generating synthetic observation and 2) computing the training objective of the neural mapping scheme. On the right side we show the evaluation principle of splitting the available satellite observations to evaluate the method on data that were not used for the inference.}
    \label{fig:method}
\end{figure}

\subsection{Neural mapping scheme}
\label{ssec:4DVarNet}

The neural mapping scheme considered for this study is the 4DVarNet framework\cite{fabletENDTOENDPHYSICSINFORMEDREPRESENTATION2021}. 
We choose this scheme due to the performance shown in the OSSE setup.
As reported in \citeA{beauchamp4DVarNetSSHEndtoendLearning2023}, it significantly outperforms the DUACS product \cite{taburetDUACSDT2018252019} in the targeted Gulf stream region. 
4DVarNet relies on a variational data assimilation formulation.
The reconstruction results from the minimization of a variational cost. This cost encapsulates a data fidelity term and a regularization term.
It exploits a prior on the space-time dynamics through a convolutional neural network inspired from \cite{fabletBilinearResidualNeural2018}, and an iterative gradient-based minimization based on a recurrent neural network as introduced by \citeA{andrychowiczLearningLearnGradient}. The overall architecture and components are similar to those presented in \citeA{beauchamp4DVarNetSSHEndtoendLearning2023}. We adapt some implementation details based on cross-validation experiments to improve the performance and reduce the training time. We refer the reader to the code for more details \cite{febvreCodeDataRelease2023a}.

\subsection{SSH Data}
\label{ssec:data}

\begin{table}[h]

\begin{tabular}{ll||cccc}
\toprule
{} & {}& Resolution & Reanalysis & Tide & DAC  \\
\midrule
NATL60 &\cite{ajayiSpatialTemporalVariability2020}               &      1/60$^\circ$ &               No &            No &                   No  \\
eNATL60-t &\cite{brodeauOceannextENATL60Material2020}         &      1/60$^\circ$ &               No &           Yes &                  Yes  \\
eNATL60-0 &\cite{brodeauOceannextENATL60Material2020}         &      1/60$^\circ$ &               No &            No &                  Yes  \\
GLORYS12-r &\cite{jean-michelCopernicusGlobal122021} &      1/12$^\circ$ &              Yes &            No &                   No  \\
GLORYS12-f &\cite{jean-michelCopernicusGlobal122021}   &      1/12$^\circ$ &               No &            No &                   No  \\
ORCA025& \cite{bernardImpactPartialSteps2006}             &       1/4$^\circ$ &               No &            No &                   No  \\
\bottomrule
\end{tabular}
\caption{\textbf{Summary table of the different synthetic SSH fields used for training}. The last column indicate whether the Dynamic Atmospheric Correction was applied on the synthetic SSH. It justify the presence of both eNATL60-0 and NATL60 to isolate the impacts of resolution and tide.}
\label{tab:data}
\end{table}

We use numerical simulations of ocean general circulation models (OGCM) to build our reference SSH datasets. Such simulations involve a multitude of decisions that affect the resulting simulated SSH. Here we consider NEMO (Nucleus for European Modelling of the Ocean) \cite{gurvanNEMOOceanEngine2022} which is among the state-of-the art OGCM in operational oceanography \cite{ajayiSpatialTemporalVariability2020} as well as in climate studies \cite{voldoireCNRMCM5GlobalClimate2013}. The selected SSH datasets reported in Table \ref{tab:data} focus on three main aspects: the added-value of high-resolution eddy-rich simulations, the impact of reanalysis datasets and the relevance of tide-resolving simulations.


In order to evaluate the impact of eddy-rich simulations, we consider NATL60, GLORYS12-f and ORCA025 free runs, respectively with a horizontal grid resolution of 1/60$^\circ$, 1/12$^\circ$, and 1/4$^\circ$. Finer grids allow for more processes to be simulated. We therefore expect higher-resolution simulations to exhibit structures closer to the real ocean and the associated trained deep learning model to perform better.
Regarding the impact of reanalysis data, we compare numerical experiments with the GLORYS12-r reanalysis and the associated free run GLORYS12-f. This reanalysis dataset relies on the assimilation of temperature, sea level and sea ice concentration observations. 
Besides, the recent eNATL60 twin simulations eNATL60-t and eNATL60-0 allow us to evaluate the impact of tide-resolving simulations.
We summarize in Table \ref{tab:data} the characteristics of the different datasets. 

\subsection{OSSE-based training setup}
\label{ssec:training}
We sketch the proposed OSSE-based training setup on the left side of the Figure \ref{fig:method}.
In order to fairly evaluate the datasets' quality as a training resource, we standardize the training procedure.
We regrid all simulations to the same resolution (1/20°) and we use daily-averaged SSH fields as training targets. We generate noise-free pseudo-observations by sampling values of the daily-averaged fields corresponding to realistic orbits of a 5 altimeter-constellation. We train all models from a one-year dataset in a Gulfstream domain from (66°W, 32°N) to (54°W, 44°N) in which we keep the same two months for validation. The hyper-parameters of the model and training procedure such as the number of epoch, learning rate scheduler are the same for all the experiments. The detailed configuration can be found by the reader in the available implementation. As training objective, we combine the mean square errors for the SSH fields and the amplitude of the gradients as well as a regularization loss for the prior model.

\subsection{OSE-based evaluation setup}
\label{ssec:eval}
As sketched on the right side of the Figure \ref{fig:method},
the evaluation setup relies on real altimetry data from the constellation of 6 satellites from 2017 (SARAL/Altika, Jason 2, Jason 3, Sentinel 3A, Haiyang-2A and Cryosat-2 ).
We apply the standardized setup presented in a data-challenge \url{https://github.com/ocean-data-challenges/2021a_SSH_mapping_OSE}.
We use the data from the first five satellites as inputs for the mapping and the last one (Cryosat-2) for computing the performance metrics. We compute these metrics in the along-track geometry. 
The evaluation domain spans from (65°W, 33°N) to (55°W, 43°N)  and the evaluation period from January 1$^{st}$ to December 31$^{st}$ 2017.  Given $\eta_{c2}$ and $\hat{\eta}$ the measured SSH and the reconstructed SSH respectively, we compute the following two metrics:
\begin{itemize}
    \item $\mu_{ssh}$ is a score based on the normalized root mean squared (nRMSE) error  computed as $1 - \displaystyle\frac{RMS(\hat{\eta} - \eta_{c2})}{RMS(\eta_{c2})}$
    \item $\lambda_x$ is the wavelength at which the power spectrum density (PSD) score  $1 - \displaystyle \frac{PSD(\hat{\eta} - \eta_{c2})}{PSD(\eta_{c2})}$ crosses the $0.5$ threshold, which characterize the scales resolved by the reconstruction (the error below that wavelength makes up for more than half of the total signal)
\end{itemize}

In Table \ref{tab:res}, we also consider the root mean square error (RMSE) as well as the nRMSE score of the sea level anomaly $\mu_{sla}$ obtained by subtracting the mean dynamic topography to the SSH. Lastly, we assess the performance degradation resulting from the transition from simulated to real data by quantifying the improvement relative to DUACS in the resolved scale $\lambda_x$ on our OSE setup as well as on
the OSSE benchmarking setup proposed in \citeA{guillouMappingAltimetryForthcoming2021}. This benchmarking setup relies on NATL60-CJM165 OSSE dataset. We refer the reader to \url{https://github.com/ocean-data-challenges/2020a_SSH_mapping\_NATL60} for a detailed description of this experimental setup.

\begin{figure}[H]
\small

\begin{center}

\includegraphics[width=\linewidth]{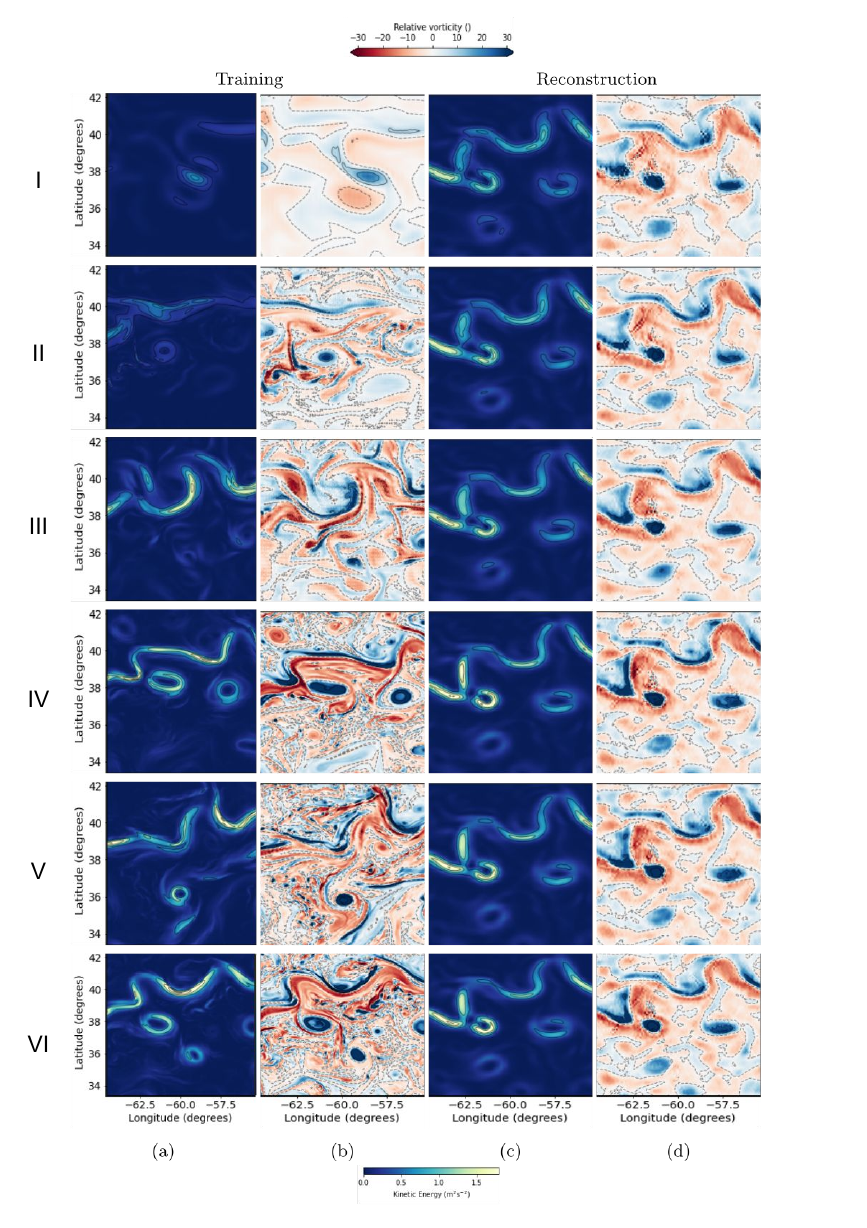}

\caption{\textbf{Samples 
Kinetic energy and relative vorticity of the training and reconstruction data of January 6$^th$ }.
The reconstructed year is 2017 while the training year vary depending on the simulation.
The first two columns (a) and (b) show the training data while columns (c) and (d) show the associated 4DVarNet reconstruction.
The kinetic energy is displayed in columns ((a) and (c)) and the relative vorticity normalized by the local Coriolis parameter in columns ((b) and (d)).
Each row shows the experiment using respectively: ORCA025 (I), GLORYS12-f (II), GLORYS12-r (III), NATL60 (IV), eNATL60-t (V) and eNATL60-0 (VI)}
\vspace{-5mm}
\label{fig:maps}
\end{center}
\end{figure}

\section{Results}
\label{sec:results}

This section details our numerical experiments for the considered real altimetry case-study for a Gulf Stream region as described in Section \ref{ssec:eval}. We first report the benchmarking experiments to assess the performance of the proposed learning-based strategy with respect to (w.r.t.) state-of-the-art mapping schemes. We then analyse how the characteristics of the training datasets drive the mapping performance. 

\subsection{Benchmarking against the state of the art}
\label{ssec:benchmarks}

We report in Table \ref{tab:bench} the performance metrics of state-of-the-art approaches including both operational observation products \cite{taburetDUACSDT2018252019,ubelmannReconstructingOceanSurface2021}, deep-learning-based schemes trained on observation data \cite{archambaultMultimodalUnsupervisedSpatioTemporal2023,martinSynthesizingSeaSurface2023} as well as methods using explicitly a model-based prior on sea surface dynamics \cite{guillouMappingAltimetryForthcoming2021,ballarottaDynamicMappingAlongTrack2020,jean-michelCopernicusGlobal122021}. We compare those methods with a 4DVarNet trained on eNATL60-0 OSSE dataset. The latter outperforms all other methods on the two metrics considered (22\% improvement in RMSE w.r.t. the DUACS product and 33\% improvement in the resolved scale). We report a significantly worse performance for GLORYS12 reanalysis. This illustrates the challenge of combining large ocean general circulation models and observation data for the mapping of the SSH.

The last column indicates that the 4DVarNet scheme leads to the best mapping scores for both the OSE and OSSE setups. For the latter, the reported improvement of 47\% is twice greater than the second best at 22\%. The performance of the 4DVarNet drops by 11\% when considering the former. By contrast, other methods do not show such differences between the OSE and OSSE case-studies. This suggests that the finer-scale structures that are well reconstructed in the OSSE setup are not as beneficial in the OSE setup. While one could question the representativeness of the OSSE datasets for the fine-scale patterns in the true ocean, real nadir altimetry data may also involve multiple processes which could impede the reconstruction and evaluation of horizontal scales below 100km.

\begin{table}[h]
\hspace{-10mm}\begin{tabular}{l||llll|rrrc}
\toprule
 & SSH  & Deep  & Calibrated on  & Physical  & rmse & $\mu_{ssh}$  & $\lambda_x$ & $1 - \frac{\lambda_x}{\lambda_{ref}}$ \\
 &  Only &  Learning &  data from &  Model &  (cm) &  () &  (km) & (\% ose, osse) \\
\midrule
(a) \textbf{4DVarNet} &  Yes & Yes & Simulation  & -- & \textbf{5.9}  & \textbf{0.91}  & \textbf{100} & \textbf{33}, \textbf{47} \\
(b) MUSTI & No &  Yes & Satellite  & -- & 6.3  & 0.90  & 112 & 26, 22 \\
(c) ConvLstm-SST & No &  Yes & Satellite  & -- & 6.7  & 0.90  & 108 & 28, -- \\
(d) ConvLstm &  Yes &  Yes & Satellite  & -- & 7.2  & 0.89  & 113 & 25, -- \\
(e) DYMOST&  Yes & No & Satellite  & QG & 6.7  & 0.90  & 131 & 13, 11 \\
(f) MIOST &  Yes & No & Satellite  & -- & 6.8  & 0.90  & 135 & 11, 10 \\
(g) BFN-QG &  Yes & No & Satellite  & QG & 7.6  & 0.89  & 122 & 19, 21 \\
(h) DUACS &  Yes & No & Satellite  & -- & 7.7  & 0.88  & 151 &  ~0,  0 \\
(i) GLORYS12 & No & No & Satellite  & NEMO & 15.1  & 0.77  & 241 & -60, -- \\
\bottomrule
\end{tabular}
\caption{ \textbf{SSH reconstruction performance of the benchmarked methods} (a) 4DVarNet from this study trained on eNATL60-0 (b) Archambault et al. (2023), (c and d)
ConvLstm-SST and ConvLstm from Martin et al. (2023), (e) DYMOST from Ballarotta
et al. (2020), (f) MIOST from Ubelmann et al. (2021), (g) BFN-QG from Guillou et
al. (2021), (h) DUACS from Taburet et al. (2019), (i) GLORYS12 from Lellouche et al.
(2021. The columns indicate from left to right: whether athe mapping schemes rely only on SSH data or also exploit additional data such as gap free SST products; if the method uses deep learning architectures; the data used to calibrate (or train) the mapping scheme; the numerical model of the ocean used for the mapping if any (QG stands for quasi-geostrophic); $\mu$ and $\lambda_x$ are the metrics as described in Section \ref{ssec:eval}}
\label{tab:bench}
\end{table}

\begin{figure}[H]
\small
\begin{center}
\includegraphics[width=\linewidth]{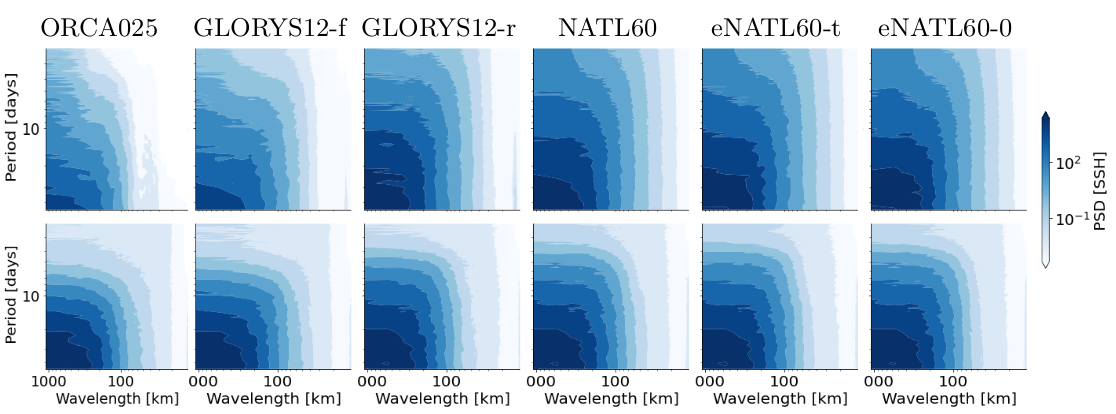}
\end{center}

\caption{
\textbf{Space-time spectral densities of the training datasets (first row) and of their associated reconstruction (second row)}. Darker blue in the lower left corner indicates higher energy at larger wavelength and periods. The different SSH fields exhibit different energy cascades when moving to  finer temporal (upward) or spatial (rightward) scales.} \vspace{-5mm}
\label{fig:spacetime_psd}
\end{figure}

\begin{table}[H]
\centering
\begin{tabular}{l||rrrrc}
\toprule
Training Data & RMSE  & $\mu_{ssh}$  & $\mu_{sla}$ & $\lambda_x$ & $1 - \frac{\lambda_x}{\lambda_{ref}}$ \\
 &  (cm) &   &   &  (km) & (\% ose, osse) \\
\midrule
NATL60 & \textbf{5.9}  & \textbf{0.91}  & \textbf{0.80}  & \textbf{98} & (\textbf{35}, --)\\
eNATL60-t & \textbf{5.9}  & \textbf{0.91}  & \textbf{0.80}  & 100 & (33, \textbf{48})\\
eNATL60-0 & \textbf{5.9}  & \textbf{0.91}  & \textbf{0.80}  & 100 & (33, 47)\\
GLORYS12-r & 6.3  & 0.90  & 0.78  & 106  & (30, 28)\\
GLORYS12-f & 6.7  & 0.90  & 0.77  & 119 & (21, 23)\\
ORCA025 & 7.1  & 0.89  & 0.76  & 126 & (17, 17)\\
\bottomrule
\end{tabular}

\caption{\textbf{Performance of 4DVarNet mapping schemes trained on different simulated datasets}. The first column shows the source of the training dataset as described in Table \ref{tab:data}; the subsequent columns indicate the reconstruction metrics described in Section \ref{ssec:eval}. Note that the NATL60 could not be evaluated on the OSSE setup since the evaluation data were used for validation during the training stage.}
\label{tab:res}
\end{table}

\subsection{Eddy-present datasets versus eddy-rich ones}
\label{ssec:resolution}

\begin{figure}[H]
\small
\includegraphics[width=\linewidth]{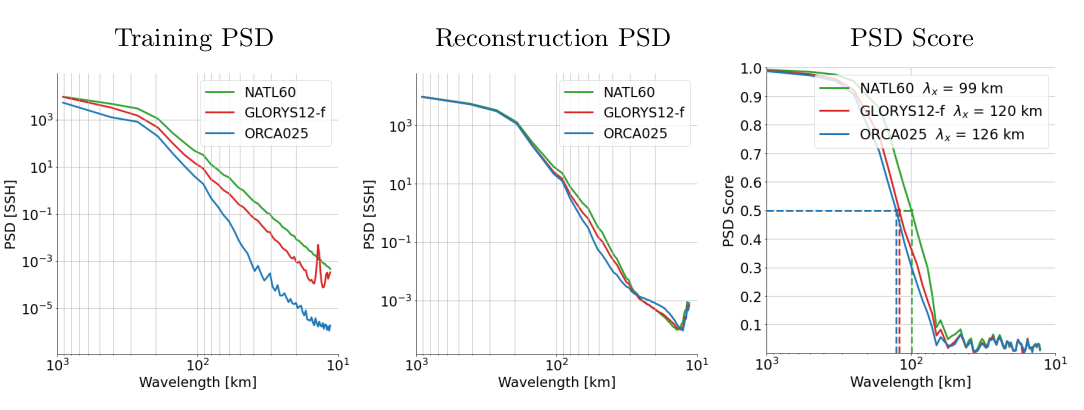}
\caption{
\textbf{Spectral analysis of the training and reconstructed SSH datasets}. We display the PSD of the training dataset (left plot), reconstructed SSH field (center plot) as well as the associated PSD score (right plot)}\vspace{-5mm}
\label{fig:respsd}
\end{figure}

We analyse here in more detail the impact of the spatial resolution of the training dataset onto the reconstruction performance. In Table \ref{tab:res}, as expected, the higher resolution grid in the ocean run simulation leads to a better mapping with a 22\% improvement in $\lambda_x$  and a 17\% improvement in the RMSE score between the experiments with the coarsest (ORCA025) and finest (NATL60) resolutions.
We also observe qualitative differences in the relative vorticity fields in Figure \ref{fig:maps}.  Residual artifacts due to the altimetry tracks appear (60°W, 39°N) for the two lower-resolution training datasets. They are greatly diminished when considering the NATL60 dataset. 
Despite these differences, the reconstructed vorticity and kinetic energy fields in Figure \ref{fig:maps} look very similar for the different 4DVarNet schemes, whatever the training datasets. By contrast, the vorticity and kinetic energy fields in the training datasets clearly depict fewer fine-scale structures and weaker gradients for the lower-resolution simulation datasets, namely ORCA025 and GLORYS12-f.
These results support the generalization skills of 4DVarNet schemes to map real altimetry tracks despite being trained on SSH sensibly different from the reconstruction. 

We draw similar conclusions from the analysis of 
the spectral densities shown in Figure \ref{fig:respsd}. The differences in the energy distribution of the training data significantly reduce in the reconstructions. 4DVarNet schemes trained from higher-resolution datasets however result in more faithful reconstruction at all scales.
The patterns observed for the temporal PSD are slightly different in Figure \ref{fig:spacetime_psd}. We do not observe the same homogenization as for the spatial PSD. Lower-resolution training datasets involve a significant drop 
of an order of magnitude for periods greater than 10 days and wavelength greater than 200km.

\subsection{Forced simulation datasets versus reanalysis ones}
\label{ssec:reanalysis}

\begin{figure}[h]
\small
\begin{center}
    
\includegraphics[width=\linewidth]{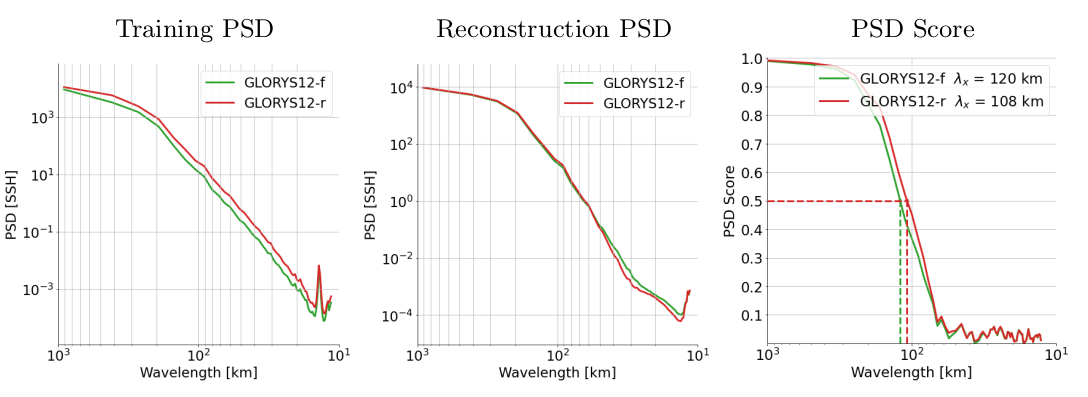}
\caption{\textbf{Spectral impact of model reanalysis}. We display the PSD of the training dataset (left plot), reconstructed SSH field (center plot) as well as the associated PSD score (right plot)}
\vspace{-5mm}
\label{fig:reapsd}
\end{center}
\end{figure}
Looking in more specifically at the effect of ocean reanalysis between the two experiments GLORYS12-f and GLORYS12-r. We can first note the impact of observation data assimilation in Figure \ref{fig:spacetime_psd} where we see how the power spectrum of the reanalysis is significantly raised compared to the free run. The spectrum is closer to ones of the higher resolution simulations. Visually we also clearly see stronger gradients in the kinetic energy in Figure \ref{fig:maps}.

We can observe a similar behavior as in Section \ref{ssec:resolution} in Figure \ref{fig:reapsd} with the gap of in spectral density being diminished between the training and reconstruction data, and the PSD score indicating a lower energy of the error at all scales for the reanalysis-based experiment.

Quantitatively in Table \ref{tab:data} we see an improvement of 11\% in both the RMSE and the scale resolved, besides training on a reanalysis increase the relative gain w.r.t. DUACS significantly more on real data (+9\%) than on simulated data (+5\%) as we can see in the right most column. This suggests that the reanalysis dataset conveys information on real world observations which improves the generalization performance.

\subsection{Tide-free datasets versus tide-resolving ones}
\label{ssec:tide}





We assess here the impact of tide-resolving simulation used as training data. We use the twin eNATL60 runs eNATL60-t and eNATL60-0. Contrary to other runs, those simulations contain barometric and wind forcing, we therefore remove the Dynamic Atmospheric Correction \cite{carrereMajorImprovementAltimetry2016} from the SSH fields. Additionally since the barotropic tide signals are removed from real altimetry tracks prior to interpolation, we also remove the signal from the training data by subtracting the spatial mean over the training domain for each hourly snapshot before calculating the daily averages.  

Given those processing steps, the two training datasets  exhibit very similar wavenumber spectra as shown in Figures \ref{fig:spacetime_psd}. 
We also find that training on those two datasets produce little differences in the reconstructions both quantitatively  (see Table \ref{tab:res}) and qualitatively (Fig. \ref{fig:maps}). The resulting performance is comparable to that of the NATL60 experiment.
 
We identify two hypotheses for explaining why tide-resolving simulation do not lead to better mapping schemes:
\begin{itemize}
    \item The preprocessing applied on the training field remove the main tide signals. We therefore effectively measure the impact of tide modeling on other ocean processes that may be less significant;
    \item The evaluation procedure applied on altimetry tracks on which the barotropic tide has been filtered may not be interpretable enough to measure the reconstruction of residual tide signals. New instruments like the KaRIN deployed in the SWOT mission may provide new ways to better quantify those effects.   
\end{itemize}

These findings provide motivation for carefully considering the purpose of the learning-based model when making decisions about the training data. In our case, explicitly modeling tide processes that are removed from the observations in the evaluation setup added overheads in the computational cost of running the simulation as well as in the preprocessing of the training data. Additionally given the considered evaluation data and metrics, we were not able to quantify any significant differences between the two trained mapping schemes.

\section{Discussion}
\label{sec:discussion}
This study has been greatly facilitated by the standardized tasks and evaluation setups proposed in data-challenges \url{https://ocean-data-challenges.github.io/}. Data-challenges are used to specify a targeted problem of interest to domain experts through datasets and relevant evaluation metrics. This preliminary work have been instrumental in constituting the comprehensive benchmark and combining methods from different teams and institution around the world. Additionally, it also constitutes a strong basis for a trans-disciplinary collaboration between the ocean and machine learning research communities.

Moreover, the results presented in this study introduce a use of ocean simulations for developing altimetry products. This opens new ways for ocean physicist, modelers and operational oceanographers to collaborate. In order to assess the range of these new synergies, it would be interesting to explore if the approach proposed here of training neural schemes using simulation data would generalize to other tasks such as forecast or sensor calibration and to other quantities like surface temperature, currents, salinity or biochemical tracers.

If the simulation-based training approach introduced here is successfully extended to other ocean problems, one could envision training large foundation deep learning models \cite{brownLanguageModelsAre} capturing the inner structure of high resolution ocean simulations which could then be used in many downstream applications. This could be the way to capitalize on all the advancement in ocean modeling without having to run OGCM numerical simulation for each downstream products.    

Furthermore, we would like to highlight the cost consideration when running numerical simulation intended for training learning based schemes. Indeed given that the eNATL60 run took 2700x CPU hours and 350x memory compared to the ORCA025 run for a smaller domain, a trade-off arises between generating multiple "cheap" trajectories versus generating a single more realistic trajectory. 

To conclude, we have shown in this study that training machine learning models on simulations datasets leads good performance on real altimetry data mapping and outperforms current state of the art approaches. The model trained on NATL60 reduces the RMSE by 18\% compared neural schemes trained on observation data and improves the scales resolved by 33\% compared to the DUACS operational product. Even the coarsest simulation considered ORCA025 provides competitive results with current operational methods. We have shown that using a more realistic SSH fields using reanalysis or higher resolution simulations increases the performances of the trained model. This is an exciting result that shows the potential for training operational products from ocean simulations and how advances in ocean modeling in operational oceanography can be beneficial. The results shown here are limited to the interpolation problem on a regional domain but the robustness of the performance shown are encouraging for further developing these results using a larger domain.

\section*{Open Research Section}
The authors provide the training data, source code, reconstructed maps and trained model for each experiments of the manuscript at https://doi.org/10.5281/zenodo.8064114.

\acknowledgments
This work was supported by ANR Projects Melody and OceaniX and CNES. It benefited from HPC and GPU resources from GENCI-IDRIS (Grant 2020-101030) and Ifremer.

\bibliography{biblio}

\end{document}